\documentclass[print, 
               notlongauthorlist 
              ]{nsr}

\volume{00}

\artnum{00}

\firstpage{1}

\datesubmitted{30 April 2025}
\datereceived{30 April 2025} 
\daterevised{19 June 2025}
\dateaccepted{19 June 2025} 
\datepublished{XX Month 2025}

\doinum{doi/number}

\copyrightyear{2025}

\author[1,2]{A. F. Zakharov\footnote{Corresponding author e-mail address: \tt alex.fed.zakharov@gmail.com}$^{\orcidlink{0000-0003-2387-6964
}}$}

\affil[1]{National Research Center -- Kurchatov Institute, Moscow, Russia}
\affil[2]{Bogoliubov Laboratory of Theoretical Physics, Joint Institute for Nuclear Research, Dubna, Russia}

\runauth{A. F. Zakharov}

\title{Shadow in the Galactic Center: Theoretical Concept -- Prediction -- Realization}

\begin{document}


\maketitle

\begin{abstract}
General Relativity (GR) was created in November 1915 and since its creation and up to now this theory has undergone many tests. The first realistic cosmological models were proposed in the works of Friedman, written in the 1920s. For a long time Friedman's cosmological works were actually banned in Soviet Union
due to philosophical reasons, since the models where the birth and evolution of the Universe occurs were considered ideologically unacceptable.
Due to great achievements in relativity and cosmology and due to increasing interest to these branches of
of science in last decades we recall a development of relativistic astrophysics and contribution of Russian researchers in these studies.
Since one of the world leaders in physical cosmology A. A. Friedman passed away in September 1925, it is reasonable to outline
the main achievements of physical cosmology over the past 100 years.
We discuss also observational and theoretical achievements in confirmations of relativistic observational  predictions for black holes, including the closest
supermassive black hole in our Galactic Center. We outline an evolution of black hole shadow from the purely theoretical concept to
observable quantities for supermassive black holes in Sgr A* and M87*. 

\end{abstract}




\section{110 years of success of GR development}

General relativity (GR) was developed by A. Einstein after intensive conversations with D.~Hilbert in November 1915
\cite{Mehra_74,Earman_78,Vizgin_79,Vizgin_01,Logunov_04}.
In spite of difficulties to create a consistent quantum gravity in numerous attempts
done by different authors \cite{Bronstein_12,Feynman_63,Feynman_64,Feynman_95,Garay_95,Kiefer_12} 
classical  GR passed all possible tests at different scales.

In 1917 a truly revolutionary event in cosmological studies has taken place since
the Universe started to be a subject for studies not only by philosophers but
also by physicists as well \cite{Luminet_08}. In this year
 A. Einstein  obtained the first cosmological model of static Universe based on his theory of relativity \cite{Einstein_17}. In this paper
Einstein assumed that the spatial distribution of matter in the Universe is uniformly isotropic and homogeneous
(now it is called the cosmological principle). Many consequent researchers used the principle in their studies after him.  
Later, A.~Eddington proved that the Einstein's solution for static Universe is instable \cite{Eddington_30},
therefore, such a solution for static Universe can not be realized in nature. 
 To construct a static Universe model A.~Einstein introduced a cosmological constant since a gravitational attraction
 must be compensated by a repulsion which is created by the cosmological constant.
 Many people repeated after G.~Gamow \cite{Gamow_1956,Gamow_70} that 
 Einstein admitted to Gamow that introducing the $\lambda$-term was the biggest Einstein's blunder, however,
 most likely it was a legend invented by Gamov \cite{Livio_13}, see also essay on the book  \cite{Simanek_13}.
 In spite of these arguments some researchers think it was a real story  \cite{ORaif_17,ORaif_18,ORaif_18a}, but
  taking into account the personalities of Gamow and Einstein, Livio's argumentation seems convincing.
 
The first realistic cosmological models were considered by Russian mathematician Alexander Friedman in 1920s \cite{Friedman_22,Friedman_24}
(English translations of these fundamental articles were published in \cite{Friedman_99,Friedman_99b}).\footnote{Interesting comments 
on these Friedmann's papers were published dedicated to jubilees of the first Friedmann's paper on cosmology \cite{Belenkiy_12,Soloviev_22,Soloviev_22b}.}
Similar results were obtained by Belgium Abbe George Lemaître \cite{Lemaitre_27} where a derivation of law which describes
velocities of distant objects as dependence of distances toward them which was observationally discovered by E.~Hubble
and later this law began to bear his name \cite{Hubble_29,Hubble_34}\footnote{In 1920s Lemaître spent a few years in USA, worked in Harvard Observatory, defended his PhD in MIT in 1927, personally knew
Vesto Slipher, who analyzed Doppler spectral shifts and found that distant nebulae are preferably  moving from us \cite{Slipher_22}. E. Hubble used the Slipher's data 
on velocities of distant nebulae and now some historians of science proposed to rename the Hubble constant \cite{MacDougal_24}.
However, we should to note that an evaluation of distance is a more complicated problem in comparison with velocity estimates using Doppler shift measurements.}. 
Later, this Lemaître's paper  was translated in English and
published in MNRAS which was the main astronomical journal in these times \cite{Lemaitre_31} (see also republication of the paper \cite{Lemaitre_13}
in the original French edition
 with comments by J. P. Luminet \cite{Luminet_13}).
Perhaps, J. Peebles was the first author \cite{Peebles_84} who noted that the Hubble law was indeed derived by Lemaître in 1927
 and Peebles emphasized the issue  at the conference talk devoted to 50 years since a creation of Big Bang cosmology
 proposed by G. Lemaître in \cite{Lemaitre_31b} (see also republication of this paper \cite{Lemaitre_11} with comments by J.-P. Luminet \cite{Luminet_11}) .

Since the derivation of the Hubble law in the English version of the Lemaître'a paper was omitted \cite{Kragh_03,Bartusiak_09,vandenBergh_11}
some people assumed 
that perhaps it was a censorship case
but after conversations with MNRAS editors M. Livio got documents \cite{Livio_11}
that Lemaître himself prepared the English translation of his original French text and
he intentionally omitted his analytical derivation of the Hubble and his estimation of the Hubble constant since
in 1929 Hubble obtained the Hubble law from observations \cite{Hubble_29}.  
D. Block proposed to apply the Stigler's law for 
the Hubble law case  \cite{Block_12}\footnote{The Stigler's law was introduced by S. Stigler and
it was motivated by the Robert K. Merton’s work on the reward system of science \cite{Merton_57}.  The simplest
form of the law is this: ''No scientific discovery is named after its original discoverer'' \cite{Stigler_80}. Very similar ideas were expressed
by V. I Arnold since he wrote ''The Arnold Principle. If a notion bears a personal name, then this name is not
the name of the discoverer. The Berry Principle. The Arnold Principle is applicable to itself'' \cite{Arnold_98}. Really as we know the Arnold's principle was formulated
earlier by Merton and Stigler.}.


At the end of 1932 Robert Milliken invited Lemaître to visit Caltech.
On January 11, 1933 Lemaître was invited to deliver a lecture at the Mount Wilson Observatory
where E.~Hubble worked. A. Einstein followed this Lemaître's lecture.
When journalists asked Einstein
about his impression on the Lemaître's cosmological model, Einstein replied ''This is
the most beautiful and satisfactory explanation of creation
to which I have ever listened!'' \cite{Lambert}.
Soon after that Duncan Aikman published interview with G. Lemaître where he claimed that there is no contradiction between
religion and science and they provide two different ways to truth \cite{Aikman_33}.
This position of Lemaître and Einstein was in contradiction with the official position of Soviet ideologists and despite the fact that Einstein was highly valued by officials of the Soviet Union, Einstein's position on philosophical issues was criticized. 
Thus, in the introduction ''On ideological vices in the book ''The Evolution of Physics'' by A. Einstein and L. Infeld'' written by S. G. Suvorov. The introduction was placed before the Russian translation by Suvorov \cite{Einstein_48} it is very fine that
the book was translated in Russian.
In the introduction Suvorov instructively informed readers about Einstein's mistakes and, in particular, noted that ''one of the judicial errors of the authors of the book is the incorrect interpretation of the development trends of modern physics''. 
Suvorov was the head of the science department in the Central Committee of the Communist Party of the Soviet Union  and later,  a deputy chief editor of the main Soviet physical journal ''Soviet Physics Uspekhi'', however, he was not a working physicist who had achieved significant scientific results but he thought (perhaps due to his high position in Soviet establishment) that he has a right to  teach one of the greatest scientists how to interpret trends in the development of physics.

In 1936 Soviet astronomer M. Eigenson\footnote{I. S. Shklovsky wrote that M. Eigenson was the communist party secretary in Pulkovo observatory in 1930s \cite{Shklovsky_91}.}  from the Main Astronomical  Observatory (Pulkovo) wrote a book \cite{Eigenson_36} where he presented the Soviet official  point of view on cosmology and this position was also expressed in an article \cite{Zelmanov_55} written by A. L. Zelmanov (in the article the Eigenson's book was cited among other important references). According to opinions of Soviet astronomers the Universe exists in infinite space and for infinite time
while expanding Universe theories proposed by Friedman, Lemaître, Gamow et al. are not allowed according to Soviet ideological point of view. It was a key difference between Soviet and Western cosmological schools \cite{Graham_87, Wetter_58}. 
It means that in contrast to other branches of physics for thirty years Soviet cosmologists were not protected by ''the nuclear shield'' against 
ignorant criticism of modern physical theories as it was discussed by V. P. Vizgin \cite{Vizgin_99}.
  The ban on the expanding models was since 1930s until 1963 when Soviet Academy of Sciences celebrated 75 years since A. A. Friedman's birthday.

It would be reasonable to recall the cases of administrative pressure on Soviet cosmologists in 1950s.
In  1955 in Soviet Academy of Sciences the  Department of Physics and Mathematics organized a special session devoted to 50th anniversary of the (special) relativity theory
where were presented cosmological talks delivered by leading Soviet theorists. 
Instructor of the Central Committee of the Communist Party of the Soviet Union (CC CPSU) A. S. Monin\footnote{Monin was  Kolmogorov's PhD student, developed
Kolmogorov's turbulent model for atmospherical diffusion, defended his PhD thesis on the subject in 1949 and thesis for the degree of the Doctor of Sciences in 1956. Monin
was the director of Oceanology Institute of Soviet Academy of Sciences for more than 20 years (from 1965 to 1987) and he was elected a full member of Russian Academy of Sciences in 2000. In 1958 A.S. Monin wrote the letter to Secretariat of the Central Committee of the Communist Party of the Soviet Union on forthcoming elections
in Academy of Sciences \url{https://ihst.ru/projects/sohist/papers/mp/1994/ilizarov.htm}. It seems unlikely that Monin did not understand the truly highest scientific level of such scientists as Landau, Tamm, Zeldovich, Lifshitz and others, therefore, apparently, the letter with such assessments of scientific reports and works of the aforementioned scientists who created glory for Soviet science was caused by opportunistic considerations.}
 and the head of the science department of CC CPSU V. A. Kirillin wrote the letter to Presidium
of Soviet Academy of Sciences where they strongly
criticised the Session where achievements of relativistic cosmology were discussed. The authors of the letter 
thought that the talks of E. M. Lifshits, Ya. B. Zeldovich, L.D. Landau and V. L. Ginzburg had significant disadvantages
and the talks were not criticised properly by the audience \cite{Blokh_05}.
At the beginning of the letter it was written ''...The program of the session was prepared by the organizing committee, which included academicians Tamm and Landau, corresponding Member of the USSR Academy of Sciences Ginzburg and prof. Lifshitz was unsatisfactory, as it included reports by Academician Landau, corresponding member of the USSR Academy of Sciences Ginzburg and professor Lifshitz,  who do not work in the field of relativity theory and are known for their nihilistic attitude to the development of methodological issues of this theory...'' Time has passed, and now we know the importance of the criticized scientists for Soviet theoretical physics, and it is hard to believe that their colleagues criticized them, since both Monin and Kirillin later became great Soviet organizers of science  and full academicians in the field of physics and its applications.

In 1940s G. Gamow proposed his version of the hot Universe model where he estimated a helium nucleosynthesis in 
primordial Universe \cite{Gamow_46,Gamow_48} and soon after that his students calculated Cosmic Microwave Background (CMB) radiation \cite{Alpher_48} which should have now temperature around $5^{\circ}$ K.  In consequent calculations CMB temperature was slightly different but always it was a several kelvins. A presence of CMB temperature was considered
as an important feature of the hot Universe model proposed by Gamow. 
To mark Einstein's 70th birthday,  the American Physical Society has published a collection of papers by famous physicists, including articles by Lemaître \cite{Lemaitre_49} and Gamow \cite{Gamow_49} 
in the special issue of {\it Reviews of Modern Physics} \url{https://journals.aps.org/rmp/issues/21/3} and Lemaître's and Gamow's papers (by both Georges) were published on adjacent pages. In 1948 a steady-state model for Universe evolution  was proposed 
by H. Bondi, T. Gold and F. Hoyle \cite{Bondi_48,Hoyle_48} in
addition to Friedmann -- Lemaître -- Gamow model discussed above. To explain the Hubble law these authors of the steady state model
assume that matter may be continuously created throughout space and time. Some outstanding astronomers (F. Hoyle, J. Narlikar, G. Burbidge and some others) supported this point of view
until the end of XX century \cite{Hoyle_00} in spite of strong criticism of their opponents. In 2014 historians of science translated Einstein's unpublished notes
written 1931 and they found that in these note Einstein considered a cosmological model where he used ideas which were very similar to ideas expressed by Bondi, Gold, Hoyle and their followers \cite{ORaif_14}. As   ''Nature'' columnist Davide  Castelvecchi noted perhaps steady-state model critics would not have been so aggressive if they had known that Einstein could once have held the same point of view \cite{Castel_14}.

Since the information from the Soviet encyclopedia was considered the absolute truth, 
information from the Soviet encyclopedia was not discussed and criticized, it was accepted that encyclopedia statements are something like a system of axioms that must be followed by any Soviet citizen (including scientists).
The article on cosmology  \cite{Zelmanov_55} clearly stated that the Friedman -- Lemaître model was unscientific and was being promoted by the West for ideological and religious reasons.
The de facto ban on the consideration of non-static Friedman -- Lemaître models of the Universe had a negative impact on the development of theoretical and observational research in the Soviet Union.
For instance, let us recall the discovery of CMB radiation in Soviet Union. 
 As it is known that the CMB radiation (which is one of the signatures of hot Universe model developed by G. Gamow in 1940s and 1950s) was discovered by T. Shmaonov at the Pulkovo Observatory several years before A. Penzias and R. Wilson (who were awarded the Nobel prize in 1978) \cite{Shmaonov_57}.\footnote{In 1940
  Andrew McKellar analysing the spectra CN and CH found that interstellar medium was very cold with a temperature of approximately 2°K,
  or more precisely 2.7°K, 2.1°K and 0.8°K \cite{McKellar_40}.
  However, astronomers did not have a correct explanation for this phenomenon for decades.}
  For instance, there is a description of this   Shmaonov's discovery in a fundamental book\cite{Peebles_09} (see also a review \cite{Trimble_06}).
 Shmaonov's supervisor was professor S. E. Khaikin who was one of the first L. I. Mandelstamm student in Moscow and he was a great expert in 
 theoretical physics,
 radiophysics, radio astronomy, and was the dean of the physics department at Moscow State University. Khaikin was fired from Moscow State University, MEPhI, and Lebedev Physics Institute after being accused of promoting idealism and Machism. A nice book \cite{Khaikin_21} was published recently on  this remarkable scientist and teacher. Shmaonov asked Khaikin about possible interpretation of his discovery and Khaikin replied that he has no explanation but
 it has to be published because it may be very important \cite{Khaikin_21}. In 1950s Soviet physicists did not cite Gamow's papers at all, Khaikin worked as the head 
 of radioastronomy department in Main Astronomical Observatory in Pulkovo, Gamow was one of the brightest representatives of Leningrad school of physics
 and there is no doubt Gamow's works were known to leading Soviet physicists, such as Khaikin, but he preferred not to declare that his student (Shmaonov) had received confirmation of Gamow's predictions. After defending his PhD thesis T. Shmaonov worked in Institute of General Physics in the laser physics laboratory headed by A. M. Prokhorov
 who got the Nobel prize in 1965 for the laser discovery and he was
 Academician-Secretary of the Department of General Physics and Astronomy (1973 -- 1993).  In 1978 when A. Penzias and R. Wilson got their Nobel prize for CMB discovery  A. M. Prokhorov criticized
 Shmaonov that he did not inform properly scientific community about his CMB studies and did not promote his discovery.

Sometimes people say, that Shmaonov's achievements were not known for many years, since they did not have an appropriate cosmological interpretation. There is a popular opinion that no one in the Soviet Union knew about the Gamow model of the hot Universe and the predictions of this model. However, this interpretation does not look correct. As mentioned earlier, in Soviet Union dynamic models of the Universe (including Gamow's models) were considered inadequate descriptions of the Universe and their consideration was not welcomed by official ideology and philosophy. In addition, despite the fact that Gamow was one of the most famous Soviet theoretical physicists, he did not return from a scientific trip abroad without the permission of the authorities. Thus, the mention of Gamow's works could be interpreted as a support for his disloyal attitude towards the Soviet government\footnote{In his interview Gamow said to Ch. Weiner on 1968 April 25 and 26 (Gamow died on August 19, 1968)
''...You see, my situation with Russian scientists is that physicists
and astronomers know that I am persona non grata, and they are afraid to write to me,
and I don't want to write to them because I bring them into trouble. But biologists don't...'' \cite{Gamow_68}.}. 

Therefore,
even if some experts understood that Shmaonov's achievements had a cosmological interpretation, they preferred not to demonstrate their understanding in order to avoid the danger of being condemned for supporting the provisions of physical cosmology, which were criticized by Soviet philosophers.
In 1963, apparently, Soviet authorities decided to reconsider the assessment of Friedman's cosmological works and the Soviet Academy of Sciences held a session of the Department of Physical and Mathematical Sciences dedicated to the 75th anniversary of Friedman's birth.
In particular, P. L. Kapitsa said in his speech \cite{Kapitsa_63}  that ''Friedman's name has so far been undeserved
oblivion. This is unfair and it needs to be fixed. We
must perpetuate this name. After all, Friedman is one of the pioneers
of Soviet physics, a scientist who made a great contribution to
domestic and world science.'' 
In July 1963, the journal "Soviet Physics -- Uspekhi" published a special issue of the journal dedicated to the 75th anniversary of A. A. Friedman, which contains articles by famous scientists such as P. Ya. Polubarinova -- Kochina, V. A. Fock, Ya. B. Zeldovich, E. M. Lifshitz and I. M. Khalatnikov, as well as Russian translations of Friedman's articles on cosmology published in German in 1922 and 1924. Moreover,  in his review  Zeldovich started to quote Gamow's papers (earlier Soviet researcher did not
mention Gamow's papers. Initially, Zeldovich criticized Gamow's papers on a hot Universe model \cite{Zeldovich_64} in spite of that Soviet
researchers did not mention Gamow and his papers after 1932 when Gamow did not return to the Soviet Union from a foreign trip,  but Zeldovich
immediately recognized a hot Universe model as a correct cosmological approach after the CMB discovery by A. Penzias and R. Wilson. Thus, it can be said that in 1963, the Soviet Union lifted the ban on discussing realistic cosmological models that consider the origin and evolution of the Universe.
To mark the centenary of Friedman's birth in 1988, the Soviet Academy of Sciences held a representative conference on gravity in Leningrad, which was attended by leading world experts in gravity and cosmology. A remarkable scientific biography was published \cite{Tropp_88} (soon after that the English translation of the book was published \cite{Tropp_93}), where, in particular, the authors wrote
''He discards a centuries-old tradition that is notoriously,
until all experience, it was considered the Universe eternal and forever motionless. He
makes a real scientific revolution. How Copernicus forced
Earth revolves around the Sun, so Friedman forced the Universe
to expand''. At this time the Soviet Union stopped to be an atheistic state and Soviet cosmology started to be a part of international
science.

Due to simplicity reasons until the end of 1990s theorists and astronomers thought that $\Lambda \approx 0$ and they adopted $\Lambda = 0$.
However, even in 1965 E. B. Gliner considered cosmological (inflationary) models with accelerating expansion
\cite{Gliner_65,Gliner_02} (see, also essay on scientific life of this remarkable scientist and personality in \cite{Yakovlev_23}). 
Such Gliner's models   looked very exotic and some outstanding Soviet cosmologists (including Ya B. Zeldovich) criticized them,
on the other hand, V. L. Ginzburg and A. D. Sakharov tried to support Gliner and his studies \cite{Yakovlev_23,Chernin_13}.
However, Gliner's results were nearly forgotten for decades  \cite{Yakovlev_23}.

In 1998 analyzing clusters with $z >0.5$ astronomers found that $\Omega_m=0.2^{+0.3}_{-0.1}$ while flat
models with $\Omega_m=1$ are ruled out by these data since probabilities for flat models are $p < 10^{-6}$ \cite{Bahcall_98} ($\Omega_m$
is a matter density in critical density units).
Since at these times cosmologists believed that the cosmological $\Lambda$ constant ''naturally'' must be vanishing an outstanding expert
 Neta Bahcall stated that the Universe has lightweight and its must expand forever \cite{Bahcall_98a}\footnote{She declared that the Universe is open
in her inaugural   article by as a new member of the National Academy of Sciences elected on April 29, 1997.
The article was published one year after her election and the achievement as interpreted one of the most significant
in astronomy and cosmology in the first part of 1998.}. 
Assuming that SNIa is a standard candle astronomers found that 
the Universe is flat ($\Omega_m+\Omega_\Lambda=1$) and $\Omega_m =0.28$ \cite{Perlmutter_98,Riess_98,Schmidt_98}. It means
that $\Omega_\Lambda \approx 0.7$ and the Universe expands with an acceleration. For this remarkable
discovery Saul Perlmutter, Adam Riess and Brian Schmidt got a Nobel prize in 2011.
Before the discovery of the vanishing $\Lambda$-term it was in the left hand side of Einstein equations
and it was treated as a geometrical part of the equations which characterizes Riemann tensor while
the expression on the right corresponds to the stress–energy–momentum content of spacetime, however, soon after the
discovery M. Turner suggested to remove the $\Lambda$-term to right hand side of the equations \cite{Turner_99,Turner_99b,Turner_99c}. Therefore, the
term could be treated as a part of the stress-energy tensor and in principle $\Lambda$ may be not a constant but a function
of spacetime and it stated to name dark energy (at this time M. Turner had a financial support from US Department of Energy in Fermi Lab and the University of Chicago). 
Consequent observations  of CMB, galaxy clusters and distant supernova data confirmed conclusions on a presence of $\Lambda$-term (dark energy).
Constraints on $\Omega_m$,  $\Omega_{\Lambda}$, $\Omega_k$ from these data are given
in \cite{Bahcall_99}.

In March 2025 DESI (Dark Energy Spectroscopic Instrument) collaboration found that it is possible that dark energy is different from an ordinary $\Lambda$ constant \cite{DESI_25}. It means that cosmological model may be different from the standard $\Lambda$CDM approach, therefore, theorists
should consider advanced cosmological models which are beyond the standard one.

\section{GR tests}

In 1942 Eisntein's assistant Peter Bergmann published one of the first books on general relativity \cite{Bergmann_42}
where he mentioned only three experiments which confirmed GR predictions, namely he discussed the Mercury anomaly, 
light deflection in observations of foreground stars near solar disk during solar eclipse  
(which were evaluated by Einstein in November 1915) and gravitational redshifts of spectral lines for white dwarfs.
In spite of the fact that observations of stars near Solar disk during solar eclipse generally confirmed
GR prediction firstly in May 29, 1919 \cite{Dyson_20}, 
accuracy of results was not very good and some scientists were rather skeptical\cite{Earman_80}. Therefore,
astronomers tried to check these results again and again with an approving precision (a detailed description of this expedition with  a discussion of
related issues was given in book \cite{Kennefick_19} published after 100 years since these remarkable observations). For instance, former Einstein's
assistant Erwin Freundlich\footnote{His attempts to measure
displacements of foreground stars to test GR predictions were described  in \cite{Hentschel_94}
(see also a nice popular book \cite{Eisenstaedt_06}).} after analysis of his observations during Solar eclipse in Sumatra on 9 May, 1929
concluded that ''(1) A deflection exists, (2) It is not Newton's, (3) It seems to be greater than Einstein's'' 
\cite{Freundlich_32}. Freundlich found that in average deflection angle is around 2".2 instead of 1".75.
However, Eddington was at the Meeting and expressed his skepticism when said ''I should like to congratulate
Professor Freundlich on the cessation of the bad luck that has dogged him for 15 years. He was the first who the first to make the attempt;
now he has at last met the success, he found interesting -- almost too interesting -- results. He has at least shown that there is  a case for further investigation. 
I find it difficult to believe that 1".75 can be wrong.
Light is a strange thing, and we must recognise that we do not know as much as about it as we thought we did in 1919, but
I should be very surprised if it is strange as all that''.

Bergmann also noted for the two first effects the theoretical magnitude of the effect only slightly exceeds the magnitude of the experimental errors
and he concluded the quantitative agreement of experiments with general relativity remains questionable.
In a more recent review and book \cite{Will_14,Will_18} C. Will presented a much more longer list for confirmations of GR predictions for different astronomical objects,
a more popular but more fresh presentation is given in \cite{Will_20}.

\section{Bright stars as test bodies to test gravity at the Galactic Center}

In order to determine the gravitational field in the area that interests us, it is necessary to follow the trajectories of the test particles. For example, Kepler, analyzing the movements of the planets in the Solar System, formulated laws that later became known as Kepler's laws, and Hooke and Newton discovered the law of universal gravitation, which described the action of gravity everywhere, and not just in the Solar System. The nearest supermassive black hole is located at the center of our Galaxy. In order to study the gravitational field in the vicinity of the center of our Galaxy and, in particular, to test the hypothesis about the presence of a black hole there, we can conduct observations of the motion of bright stars there. 
In his Nobel lecture R. Genzel outlined a way to an evidence that there is a supermassive black hole in the Galactic Center \cite{Genzel_22}.
In 1960s -- 1980s it was an intensive discussion concerning an existence of supermassive black holes and and energy release in quasars and AGNs, in particular, a presence of supermassive black hole in Galactic Center. In 1969 D. Lynden-Bell  proposed binding energy of black holes as a source of an energy release in quasars \cite{Lynden_Bell_69}. 
 Soon after that J. Bardeen noted that an energy release could be higher for Kerr black holes \cite{Bardeen_70}.
 D. Lynden-Bell and M. Rees considered different mechanisms for an energy release in the Galactic Center  \cite{Lynden_Bell_71}.
However, some researchers thought that the black hole mass in GC can not be so high, for instance,
 Soviet astrophysicist L. M. Ozernoi and his group from the Lebedev Physics Institute of Soviet Academy of Sciences
insisted that models of supermassive black hole in the Galactic Center should be ruled out \cite{Ozernoi_76,Ozernoi_79}.
However, astronomers led by Ch. Townes  (University of California (Berkeley)) analysing results of
observations for interstellar clouds concluded it has to be a compact mass around $ (2-4) \times 10^6 M_\odot$
in the Galactic Center \cite{Wollman_77,Lacy_80,Lacy_82,Serabyn_85,Crawford_85}.
British theorist M. Rees also argued that the black hole mass in the GC should be around $5 \times 10^6 M_\odot$ \cite{Rees_82},
meanwhile D. A. Allen and R. H. Sanders thought that black hole mass should be around $100~M_\odot$ \cite{Allen_86}.
Different opinions concerning the black hole mass were a good stimulus to find new arguments to convince opponents.
In 1987 R. Genzel and C. Townes presented black hole mass estimate in the range $(2-3)\times 10^6 M_\odot$
and concluded that the evidence for supermassive black hole mass ($M \approx 10^6 M_\odot$) is substantial but not convincing \cite{Genzel_87}.

Using diffraction-limited ‘speckle’ imagery starting in 1991/1992 on the 3.5m New Technology
Telescope (NTT) of the European Southern Observatory (ESO) in La Silla/Chile,  MPE astronomers
found proper motions of stars as close as ~0.1” from Sgr A* which is actually coincides with our Galactic Center (GC)\cite{Eckart_96,Eckart_97}. Since 1995
Ghez's group (University of California, Los Angeles) started to use 10-m Keck telescope to monitor the bright stars near the Galactic Center 
and the authors obtained similar results \cite{Ghez_98}. In 2002 the MPE group started to work with 8.2 m
Very Large Telescope (VLT) on Paranal and the authors declared that they found a star (S2) which was approaching Sgr A* at
10 -- 20 mas and its period was estimated as 15.2 years \cite{Schodel_02} and the black hole mass was estimated as $M(0)=3.7 \times 10^6 M_\odot$. 
Since the S2 (S02) star orbit is highly elliptical ($e=0.88$)
with a peri-distance of 14 mas (17 light hours or 1400 $R_S$ ($R_S$ is the Schwarzschild radius for supermassive black hole at GC), for $M(0)=4.26 \times 10^6 M_\odot$ according the corrected estimates\cite{Genzel_21}).
Similar eatimates for orbital parameters of  S02 star are given by the Keck group \cite{Ghez_03}.
Monitoring its trajectory gives an opportunity to test gravitational effects in GC as it was done by Keck and VLT groups.
Even preliminary results on the trajectory of S2 star gave an opportunity to constrain 
extended mass distribution  \cite{Nucita_07} and parameters of stellar cluster and dark matter distribution
near the GC \cite{Zakharov_07}.
At the ESO Workshop held in Garching,  4-8 April 2005 it was proposed to combine four VLT telescopes in an interferometer   
(General Relativity Analysis via VLT InTerferometrY,
or GRAVITY) but proceedings of the Workshop were published only after 3 years since the Workshop date \cite{Einsenhauer_08,Paumard_08}.
GRAVITY facilities were designed and made by a French-German-Portuguese Consortium of 6 Institutes (plus ESO) and installed on Paranal in July 2015
\cite{Genzel_21}. A description of GRAVITY interferometer was done in \cite{GRAVITY_17}.
Currently there are plans to extend baseline and create a VLTI interferomer  with a baseline about a few km \cite{Eisenhauer_24} where
key components will be VLT's and the GRAVITY interferometer.
In August 2012, the IAU congress was held in Beijing and the Chinese journal (Research in Astronomy and Astrophysics) published a series of reviews on the most pressing issues in astronomy by leading experts. In particular, M. R. Morris, L. Meyer and A. M. Ghez published a review   on observations of the Galactic Center in different bands and the interpretation of these observations \cite{Morris_12}.
Commenting observational results of both groups \cite{Plewa_15,Sakai_19}
R. Genzel claimed \cite{Genzel_21} that
''the gravitational potential indeed is dominated by a
point mass, whose position is identical within a mas uncertainty with that of the radio source
Sgr A*''. 
 In May 2018, the star S2 was at the apocenter of its orbit, a fairly good approximation of which is an ellipse centered at Sgr A*.
 One of the well-known tests of general relativity is the gravitational redshift of spectral lines and the effect is  strongest 
 in the strongest gravitational field which is near the apocenter.
The GRAVITY collaboration measured the gravitational redshifts and confirmed GR predictions \cite{GRAVITY_18,GRAVITY_19,GRAVITY_19b},
in 2019 the Keck group published its results on gravitational redshifts of S2 star near the apocenter \cite{Do_19}.

Due to a slow progress in studies of dark matter and dark energy puzzles, extended theories of gravity have been proposed to try to explain
these puzzles as gravitational phenomena \cite{Capozziello_08,Zakharov_09,Capozziello_11,Faraoni_11}.
Sometimes, $f(R)$-theories of gravity do not have a Newtonian limit in a weak gravitational field approximation
and Solar system data put severe constraints on these theories \cite{Zakharov_06}.
Observations of the S2 star trajectory also gave an opportunity to constrain parameters of these theories \cite{Borka_12,Zakharov_14b}.
In 2010s it became clear that the accuracy of observations of the orbits of bright stars does not yet allow us to verify the predictions of the General Relativity, but nevertheless it is possible to limit the parameters of alternative theories of gravity that have been actively discussed in recent years.
Thus, in the paper \cite{Borka_13}, restrictions on the parameters of the Yukawa theory were obtained
 and later, in the work \cite{Zakharov_16,Zakharov_16b}, restrictions on the graviton mass were obtained, and this restriction is comparable with the restrictions obtained from gravitational-wave experiments done by LIGO -- Virgo -- KAGRA (LVK) collaboration \cite{LIGO_16,LIGO_21}
 and the best mass constraint $m_g < 1.76 \times 10^{-23}~eV$ the LVK collaborations obtained from analysis of O1 -- O3a runs.  
Later, our constraints of graviton mass done from observations of bright stars near the Galactic Center were
improved by Hees et al. using new Keck group's data \cite{Hees_17,Hees_17a,Hees_17b}.
Our constraints on graviton mass together with other bounds done other observations are presented in PDG table for graviton mass bounds \cite{Navas_24}.
Further improvement of graviton mass bounds with observations of bright stars near the GC were found in \cite{Zakharov_17b,Zakharov_18a,Zakharov_18}.
In paper \cite{Zakharov_18b} it was shown that there is an opportunity to constrain a tidal charge of black hole from analysis of bright star trajectories. 
The conventional model for the Galactic Center with a supermassive black hole and some alternative models were discussed in \cite{Zakharov_18c,Zakharov_19}.
Analysing trajectories of bright stars the GRAVITY collaboration considered constraints on parameters of alternative models for GC \cite{GRAVITY_19c}.
In 2020 the GRAVITY collaboration detected the Schwarzschild precession for the S2 star orbit \cite{GRAVITY_20}
(it was the second GR effect confirming GR predictions for the Galactic Center).
Using GRAVITY results on the Schwarzschild precession parameters of extended theories of gravity were constrained in \cite{Borka_21},
while an influence of a bulk distribution of matter on the precession was discussed in \cite{Jovanovic_21}. 
 Using observational data done by GRAVITY collaboration Constraints of Yukawa gravity parameters were found in \cite{Jovanovic_23}.
Taking into account the Schwarzschild precession found by the GRAVITY collaboration
graviton mass bounds were found in \cite{Jovanovic_24Symm,Jovanovic_24}.
Recently new bounds on the fifth force parameters were discussed from observational data obtained by the GRAVITY collaboration \cite{GRAVITY_25}.

It is generally believed that the Galactic center contains a supermassive black hole surrounded by a  small mass spatial distribution of matter.
However, Ruffini, {Arg{\"u}elles} and {Rueda} proposed replacing the supermassive black hole with fairly dense dark matter \cite{Ruffini_15}
(later this approach started to call the RAR-model). In the simplest approximation we have that a dark matter density in core is a constant.
A few years ago, Becerra-Vergara et al. claimed that the RAR-model provided a better fit for trajectories of bright stars in the Galactic Center 
\cite{Becerra_21}. Really, as it was shown in papers \cite{Zakharov_2022MNRAS,Zakharov_2022MUPB}
in the framework of the RAR-model stars move along elliptical trajectories since there is a harmonic oscillator potential 
 inside a ball with a constant density and in this model  the centers of these ellipses coincide with the origin of coordinates (or, in other words, with the center of a sphere of constant density), while observations done by the Keck group and GRAVITY collaborations showed that, in the first approximation, the stars move in a Newtonian potential and the center of symmetry of the gravitational potential coincides with the focus of the ellipse, therefore,  a ball with a constant density can not be
 a suitable substitution for supermassive black hole in the Galactic Center. 

Analyzing orbits of bright stars the GRAVITY collaboration obtained constraints on bulk matter distributions inside orbits of bound bright stars
\cite{GRAVITY_2022AA,GRAVITY_2024AA,GRAVITY_2024MNRAS}. A comprehensive review of observational tools for studying black holes is presented in the article
\cite{Genzel_24}.

\section{Black hole shadow in brief}

In 1973, James Maxwell Bardeen discussed a thought experiment that suggested that there was a glowing screen behind a black hole. In this case, the observer will see a small dark spot (shadow) on the background of the screen \cite{Bardeen_73}. 
It is a purely GR effect since there are no shadows for massive point in Newtonian gravity.  
For a long time, this model was not applied to astrophysical black holes, since astronomy does not have a bright screen behind a black hole, and for a long time in our Galaxy, stellar-mass black holes were mainly considered, whose shadow size is about a million times smaller than that of the shadow itself, the size in the center of the Galaxy (GC). In addition, questions were raised about the possibility of distinguishing a shadow from a dim astronomical object. Using estimates of the expected parameters of interferometers being created in the near future and updated about 20 years ago, which turned out to be fairly accurate estimates of the mass of the black hole in GC, Zakharov et al. \cite{Zakharov_05a} predicted the possibility of reconstructing the shadow in the center of the Galaxy using a ground-based or space-based interferometer operating in the millimeter or submillimeter range (also in this work for the first time the capabilities of the Millimetron interferometer for such needs were mentioned). Our prediction became true in May 2022, when the Event Horizon Telescope (EHT) collaboration presented the results of reconstructing the shadow of the center of our Galaxy (the shadow of the supermassive black hole in M87 was reconstructed in 2019). These reconstructions were based on EHT observations conducted in 2017, so the analysis of observational data for M87* was carried out for more than 2 years, and for Sgr A* for more than 5 years.  
For the Reissner-Nordstrom metric the authors \cite{Zakharov_05b} obtained analytical expressions for the shadow size as a function of charge, and later these results were generalized to the case of tidal charge \cite{Zakharov_12,Zakharov_14}.
These results were presented and promoted at different conferences \cite{Zakharov_2005dmap,Zakharov_protvino04,Zakharov_Texas_05,Zakharov_2005SIGRAV}.
   We were discussing the possibilities of estimating the parameters of alternative theories of gravity using shadow size estimates made by the EHT collaboration, in particular, based on these observations, the tidal charge can be estimated \cite{Zakharov_22,Zakharov_22AAPT}. We are also discussing the possibilities of using Millimetron equipment for shadow reconstruction in the M87* and Sgr A*. In our recent study \cite{Zakharov_25,Zakharov_25b,Zakharov_25c}, we discussed the formation of shadows in cases where naked singularities or wormholes replace black holes in the centers of galaxies and considered the cases where generalizations
of Kerr black hole metrics have no shadows.

\section{VLBI ideas}

Ideas of VLBI (Very-Long-Baseline Interferometry) were proposed by Leonid Ivanovich Matveenko in beginning of 1960s. 
Matveenko recalled the initial stage of development of this concept in \cite{Matveenko_2007R,Matveenko_07,Matveenko_19}.
After 10 years since the start of development of this concept L.~I.~Matveenko wrote an article with the title
''A radio telescope the size of a Globe'' \cite{Matveenko_73} in one of the most widely read popular science magazine "Science and Life" in Soviet Union (for example, the amount of copies of issue 10 in 1973 was 2,900 thousand copies). 
Really, science was very popular in USSR at these times.
This slogan was so nice that many people used it after Matveenko
usually forgetting to specify the author who came up with and used this slogan for the first time.

According to Matveenko's memoirs
in 1960s Matveenko worked as a radio physicist at the Deep Space Network (DSN) Center near  Evpatorija  (Crimea). The Center was established for control and communicate with Soviet spacecrafts. The Center had a radio interferometer with baseline length around 500~m.
 In 1962 Matveenko decided to increase a baseline and proposed the first interferometric experiment using telescopes DSN and near Simferopol. In autumn 1962
 Matveenko reported about his proposal at the seminar of radio laboratory (head V.~V.~Vitkevich) in Pushchino
 and this proposal was not supported  \cite{Matveenko_2007R,Matveenko_07}. Moreover, ''V.~V.~Vitkevich excused that the Laboratory cannot
support publishing the paper or carrying out an experiment'' \cite{Matveenko_07}. Matveenko requested a support from radio astronomers G. B. Sholomitskii and  N.S.
Kardashev
 in Sternberg State Astronomical 
Institute (SSAI) of Moscow State University and reported these ideas at the institute seminar  \cite{Matveenko_2015}. The institute director D. Ya. Martynov supported
 these ideas and recommended to take a patent. The head of radio astronomy department in SSAI was I. S. Shklovsly.  In
 1962 SSAI\footnote{In his notes on  development of radio astronomy in Soviet Union I.~ S.~Shklovsky noted an importance of VLBI techniques \cite{Shklovsky_82,Shklovsky_96}, developed by   N.~S.~Kardashev,  L.~I.~Matveenko  and
G.~B.~Sholomitskii, in these notes Shklovsky changed the order of the authors was in the patent application and in Soviet Radiophysics journal publication
\cite{Matveenko_65}. Shklovsky emphasized that this techniques are especially useful
   for observations of compact objects, for instance quasars, however,
 the method of VLBI observations was not listed among significant achievements of the radio astronomy department in  SSAI \cite{Gindilis}. } 
 sent  the VLBI proposal by L.~I.~Matveenko, N.~S.~Kardashev, and
G.~B.~Sholomitskii in the Soviet Patent Bureau. However, experts in the Patent Bureau did not approve the proposal and
the noted that usually they issued patent for scientific or technical result but not method \cite{Matveenko_07}. 
Matveenko requested the Patent Bureau to give a patent or a permission for publication in journal and finally
the paper was published \cite{Matveenko_65} almost after 2.5 years after the first application for patent. 
However, the authors' proposal on the possibility of using ground-based space interferometry was removed from the article at the request of the reviewer. Subsequently, this idea was given in a popular article \cite{Matveenko_73} (on the 20th anniversary of the idea, Matveenko's article was published in a popular journal \cite{Matveenko_82}).

In summer 1963 the director of    Jodrell
Bank Radio Observatory Sir Bernard Lovell visited Soviet Union as a special guest of Soviet Academy of Sciences (at these times the
Observatory had the largest radio telescope in world and the telescope could track the first Soviet Sputnik \cite{Spinardi_06}).
Shklovsky invited  Matveenko to present his ideas on long range interferometry in front of the distinguished guest\footnote{Lovell wrote
in his diaries that he felt some kind of sickness and he was unhappy after this visit, see
\url{https://luna.manchester.ac.uk/luna/servlet/detail/Manchester~14~14~1623~192985?qvq=q\%3Abernard+lovell&mi=1&trs=5}. Perhaps due to these circumstances
Soviet -- British collaboration was not started as previously agreed in the Crimea in the summer of 1963 and it was written in  a Memorandum
concerning joint observations with interferometer between radio telescopes in Evpatorija and Jodrell
Bank at $\lambda = 32~cm$ \cite{Matveenko_2007R}.}. Lovell approved the Matveenko's idea, but expressed doubt that there are powerful, practically point-like astronomical sources for the observation of which such an angular resolution is necessary. However, we have to remind that in 1963 quasars were discovered
which are very compact an bright and VLBI facilities started to be very useful for observations of these objects.

Reminding the VLBI idea realization in Soviet Union L. I. Matveenko wrote:
''Without going into the details of the formation and development of the VLBI, I can only note that the main problems in our country were not bureaucratic obstacles, but the ''support'' of colleagues \cite{Matveenko_2007R}''.
Mathematician M. I. Monastyrsky (ITEP) expressed a  very similar idea analyzing several cases of outstanding Soviet mathematicians \cite{Monastyrsky_09}:
''By the way, we (in Russia) love to lament about the underestimation of Russian scientists in the West. But a careful analysis of the real facts shows that the greatest obstacles to the recognition of Russian (Soviet) scientists are other Russian (Soviet) scientists.''\footnote{Here we could remind A. Friedman 
whose works were considered relevant in the USSR, Gamow, whose works Soviet physicists preferred not to cite for reasons known to them, Matveenko, whose ideas were not supported by the leadership, Gliner, whose ideas were not supported by many influential physicists, despite the support of such leading theorists as Ginzburg and Sakharov.}

In 1976 the first global telescope was created, or more precisely, on April 26 and May 6, 1976, radio telescopes in the European USSR, the United States, and Australia were linked together in order to observe ${\rm H}_2$O maser sources with an angular resolution of 0.1 milliarcsec \cite{Batchelor_76}.
Four antennas formed this global telescope, namely, 22-m (Simeiz), 26-m Maryland Point Observatory of Naval research Laboratory,
40-m Owens Valley Radio Observatory in California, 64-m NASA telescope in Tidbinbilla (Tidb) in Australia. 

In the 1970s, Matveenko participated in interferometric observations of compact radio sources (3C 84, 3C 273, 3C 345, NRAO 150) at $\lambda = 1.35~$cm, 
where radio astronomers used the following telescopes: 22-m Simeiz (USSR), 20-m Onsala (Sweden), 100-m Effelsberg (Germany), 37-m Haystack (Westford, Massachusetts, USA)
\cite{Pauliny_78}.\footnote{According to NASA ADS L. I. Matveenko had eight joint papers with future Nobel prize winner R. Genzel in small groups of co-authors, see, for instance
results of maser line observations with the Crimea -- Effelsberg and Haystack -- Green Bank observations in November 1979 \cite{Matveenko_82b}.}
In 1976 the Crimea -- Haystack interferometer observed the source W51 with maser lines at OH and ${\rm H}_2$O at $\lambda = 1.35~$cm \cite{Matveenko_78}.  

In 1980s L. I. Matveenko proposed a project with space -- ground interferometry \cite{Kostenko_82}. Later, the Japanese HALCA-VSOP and Russian Radioastron missions were
launched and operated on the orbit. However, after processing the results of observations obtained with the help of these space radio telescopes, it seems impossible to assert that fundamental discoveries have been made that could be considered appropriate to the space cost of projects.

\section{Synchrotron radiation in astrophysics}

The Institute of Theoretical and Experimental Physics in Moscow (ITEP) was founded by academician A. I. Alikhanov in December 1945,
and the scientific community of the institute could have celebrated the 80th anniversary of its founding, but since 2022 the institute has ceased to exist.
The head of theoretical department in ITEP in 1950s–1960s was a famous Soviet theorist I.Ya. Pomeranchuk. Brief essays
on the Pomeranchuk life and his scientific achievements are presented in \cite{Okun_03, Berestetskii_67}. He made a number
of remarkable discoveries, in particular,  he predicted an existence of electromagnetic emission from electrons moving in
magnetic fields  \cite{Pomeranchuk_40,Iwanenko_44,Artsimovich_45} (this physical phenomenon
was analyzed earlier by G. A. Schott \cite{Schott_12}, however,
in 1940s his studies were nearly forgotten). 
Since in 1940s rather powerful accelerators were in action \cite{Kerst_42}
soon after publications
of the papers by Pomeranchuk, Iwanenko
and Artsimovich the emission was discovered at a
synchrotron and it is started to name it a synchrotron
emission. Really, the first detection of X-ray radiation
from electrons in 70-MeV synchrotron has been done
by Elder et al. \cite{Elder_47,Elder_47b}. The history of the discovery
is presented in \cite{Blewett_46,Baldwin_75,Pollock_83}. According to the I. I. Gurevich’s
opinion this Pomeranchuk’s work on magnetic
bremsstrahlung (synchrotron) radiation should have
been crowned with a Nobel Prize \cite{Gurevich_88} (I. I. Gurevich was an outstanding Soviet physicist
and he made a significant contribution to the implementation of the Soviet atomic project).

One of the simplest explanation for origin of synchrotron emission is given in Chapter 34 in the nice classical textbook (see, for instance, reprinted edition \cite{Feynman_10}). 
A brief history of discovery and development of synchrotron radiation was presented in  \cite{Blewett_98} (and it was repeated in significant points in \cite{Margaritondo_22}).
In these papers
the authors noted that a significant progress in understanding the synchrotron radiation  was done after
publications of papers by Alfred Lienard \cite{Lienard_98} and Emil Wiechert \cite{Wiechert_01}, where the authors introduced delayed potentials.
Different applications of synchrotron radiation in science and technology are outlined in \cite{Margaritondo_17}.
 The first particle accelerator Tantalus fully dedicated to synchrotron light
experiments started to operate in 1968 in Wisconsin, experiments which were done in the first years of its operation are outlined
in \cite{Margaritondo_08}. A general description of such facilities of the first generation was given in \cite{Godwin_69}.

In the mid of 1940s I. S. Shklovsky followed a talk delivered by Yu. B. Kobzarev. Kobzarev reported that radio physicists detected
radio emission from Sun. Shklovsky asked himself and after analysis of the problem he concluded that radio emission could be generated by electrons in moving magnetic fields in Solar corona \cite{Shklovsky_46}. V. L. Ginzburg obtained very similar conclusion in his first astronomical paper \cite{Ginzburg_46}.
On May 20, 1947, a total solar eclipse was to occur in Brazil, and the Soviet Academy of Science decided to send an expedition to observe the solar eclipse in the radio and optical ranges. Papaleksi was appointed as the leader of the expedition.
Papaleksi believed that observing the Sun during a solar eclipse would allow him to test theoretical predictions that radio emission was generated in the corona.
But in February 1947, Papaleksi died and S. E. Khaikin was appointed as the leader of the expedition. Two famous theorists  (Shklovsky and Ginzburg)
predicted an existence radio emission from Solar corona participated in this expedition. Both of them left their memoirs about this exotic trip
\cite{Ginzburg_17,Shklovsky_91}. 
Khaikin and Chikhachev performed this delicate experiment, in which not a telescope, but a ship tracked the position of the eclipsed Sun, and confirmed the predictions of theorists \cite{Khaikin_21}. After that, using their radio observations
of Sun during Solar eclipse at the wavelength $\lambda = 1.5$~m
 Khaikin and Chikhachev published papers where they described their discovery of radio emission from Solar corona
\cite{Khaikin_47,Khaikin_47b}.
 Only many years later this discovery was officially registered and the Committee on Discoveries and Exploration under the Council of Ministers of the USSR was registered in the State Register of the USSR on April 28, 1970 under No. 81 with priority from October 28, 1947 discoveries by Khaikin S.E., Chikhachev B.M., Papaleksi N.D. and issued a diploma on September 14, 1971 on the discovery of radio emission from the solar corona \cite{Khaikin_21} (it would be reasonable to remind that  Papaleksi N.D passed away 
 on February 3, 1947, Khaikin  died on July 30, 1968, therefore, these two remarkable scientists did see an official state recognition
  of their remarkable discovery). Optical observations of stars near Solar disk during the total solar eclipse to test
  GR predictions were not done due to bad weather conditions.

The Crab nebula was discovered around two hundred years and in 1950s  there were many observations but a central engine mechanism was unknown.
and in 1953 a famous Soviet astrophysicist I.S. Shklovsky proposed a synchrotron emission as a
key mechanism to explain observational data in a wide range of frequency band from radio to optics  \cite{Shklovsky_53}.
Later the Shklovsky’s model was confirmed with consequent X-ray data. In his memoirs Shklovsky noted
the idea to use a synchrotron mechanism for the Crab nebula was among of his brightest insights \cite{Shklovsky_82,Shklovsky_96}.
Now we know that synchrotron radiation plays a key role in generation of electromagnetic radiation near many different astronomical objects, including
pulsars and black holes.

\section{Shadows in M87* and Sgr A*}

As observational and experimental technologies advance, the possibilities for testing theoretical predictions of various theories, including Einstein's theory of relativity, increase. However, theorists are faced with important questions related to which objects should be observed, when, and with what observational facilities.
As it was noted earlier, Matveenko 
reported his VLBI  ideas  to a famous British radio astronomer Lovell in 1963, 
he expressed skepticism regarding the need to built  such facilities, since powerful compact radio sources were unknown, however, as Shklovsky noted, quasars, which are such sources, became known already in 1963.
It was also mentioned that the first space -- ground interferomer was proposed by L. I. Matveenko and he promoted this idea in 1980s and
it was realized only by Japanese HALCA (Highly Advanced Laboratory for Communications and Astronomy), also known for the project name VSOP (VLBI Space Observatory Programme)
which was launched in 1997 and officially stopped in 2007. In 2011 Russian space -- ground Radioastron mission started and it was stopped to operate in 2019.
In spite of significant amount of observations with these facilities breakthrough results were not obtained.
Japanese Space Agency cancelled the VSOP successor ASTRO-G (VSOP-2) due to the risk of not achieving significant scientific results. 

In 2000s when a scientific program of Radioastron was in preparation it was know that a black hole mass in the Galactic Center is around $4*10^{6}M_\odot$ 
and distance is around 8~kpc, the Schwarzschild radius size is around $10~\mu$as, while the angular resolution
of ground-- space telescope was expected to around $8~\mu$as at the shortest wavelength $\lambda =1.35$~cm.
Since angular resolution of such facilities was comparable with a typical scale of the GR effects it would be reasonable to expect
to find an interesting GR effect in observations of GC with Radioastron facilities.
In 2000 Falcke et al. considered a toy model where they simulated an opportunity to observe shadow around the black hole at GC
in different wave band \cite{Falcke_00}. The authors concluded that there is a chance to observe a shadow at  $\lambda =0.6$~mm  
while interstellar scattering practically destroyed shadows for $\lambda =1.3$~mm. Using available estimates for the black hole mass $2.6 \times 10^{6} M_\odot$ \cite{Eckart_96,Ghez_98}
Falcke et al. predicted the angular diameter of the shadow in Sgr A* from the GR calculations alone to be around $(30 \pm 7) \mu as$ \cite{Falcke_00}.
An angular resolution of VLBI interferometer as Event Horizon Telescope has now is around $25~\mu$as, therefore, perspectives
to observe shadow for the Galactic Center did not look very optimistic and Falcke et al. mentioned that perhaps it would necessary to use
wave length below $0.2$~mm or even to use MAXIM interferometer to solve such a problem \cite{Falcke_00} (developed by W. Cash, see detailed \cite{Cash_00,White_00}
but unfortunately, the MAXIM project was cancelled around two decades ago due to high cost and an absence of guaranteed sound results).

Due to a presence of secondary images near black hole shadows \cite{Holz_02} we predicted
that black hole shadow at the Galactic Center could be reconstructed from 
VLBI observations in mm and sub-mm bands and in this case ground -- space observations with Millimetron facilities
will be very  useful \cite{Zakharov_05a}.
  In April 2017 EHT collaboration with a global VLBI network observed M87* and Sgr A* at 1.3~mm wave length.
  In April 2019
  the first publications on  shadow reconstructions for M87* were done
  \cite{EHT_17_p1}. The authors
  adopted    a  distance toward this object  as $(16.8 \pm 0.8)$~Mpc
   evaluated a shadow diameter $(42 \pm 3) \mu$as and 
found that the black hole mass $M = (6.5 \pm 0.7) \times 10^9 M_\odot$. The authors declared 
that ''the photons at 1.3mm wavelength observed by the
EHT are believed to be produced by synchrotron emission'' \cite{EHT_17_p1}.  Later, the EHT collaboration
reported polarization distribution in M87* ring around the shadow  \cite{EHT_21_p1}
and magnetic field map distribution which is consistent with polarization \cite{EHT_21_p2}. The
data strongly supported an initial assumptions of the model that astronomers
observed the ring due to synchrotron emission of electrons moving in magnetic fields around the black hole in M87*.

In May 2022  the EHT collaboration presented a shadow reconstruction for Sgr A* \cite{EHT_22_GC_p1}
after more than 5 years of data processing after observations in April, 2017. 
The authors declared that the shadow diameter is $51.8 \pm 2.3~\mu$as (with 68\% credible interval).
This fact illustrates difficulties in the problem of shadow reconstruction
for object with strong variability on a scale of tens of minutes. However, this excellent result
was a confirmation of our prediction done in 2005 \cite{Zakharov_05a}, that the black hole shadow for GC can be reconstructed
from VLBI observations in mm band and shadow diameter should be around 50~$\mu$as
(see, also discussions of the issue in \cite{DeLaurentis_23,Bambhaniya_24}).

In 2024 the EHT collaboration reported polarization distribution in the ring near the black hole shadow in Sgr A* \cite{EHT_24_GC_p1}
and distribution of magnetic fields near the horizon of the black hole \cite{EHT_24_GC_p2}. These observational data support
 a synchrotron radiation model for detected flux of electromagnetic radiation from the Galactic Center.

Thus, the concept of a black hole shadow has gone from a theoretical concept proposed by J. Bardeen \cite{Bardeen_73} to the prediction that the size and shape of the shadow of a black hole in the Galactic center can be obtained from VLBI observations in the mm or submillimeter range  \cite{Zakharov_05a} and the reconstruction of the shadow by the collaboration in 2022 from their observations in April 2017 \cite{EHT_22_GC_p1}.

\section{Are shadows 100\% confirmations for presence of SMBHs in M87* and Sgr A*?}

In popular literature and elsewhere, one can find the assertion that observation of the shadow of black holes and analysis of the trajectories of bright stars provide 100\% evidence of the existence of black holes in these objects. If it is necessary to choose a suitable model from a small number of alternatives under consideration, then the galactic center model may be preferable to other models.
Therefore, in mathematical terms, when explaining a natural phenomenon, proof of the uniqueness of the model used is usually not given, since one can usually only speak of the preference of one model over another. Below we give an example of a metric that differs from the Schwarzschild black hole metric, but which similarly describes the shadows of black holes. 

It is known that the influence of spin on the deformation of the shadow of a black hole decreases as the position of a distant observer approaches the axis of rotation of the black hole and the position of the observer for the cases of M87* and Sgr A* far from the equatorial we will limit ourselves to the case of spherical symmetry of the metric, which will simplify our subsequent reasoning. Let's consider a class  spherically symmetric static metrics. Ideas of our analysis is very similar to Friedmann's
consideration done in his first cosmological paper \cite{Friedman_22,Friedman_99}.

Let us recall the expression for such a metric
\begin {equation}
ds^{2}=A(r)dt^{2}-A^{-1}(r)dr^{2}+
r^{2}(d{\theta}^{2}+{\sin}^{2}\theta d{\phi}^{2}),
\label{M_0}
\end {equation}
Below we measure linear distances $r$ in $M$ ( $M$ is a mass of a compact object)
From the spherical symmetry of the metric we have that the motion occurs in one plane ($\theta=\pi/2$) and the angular momentum is conserved
$r^2 \dfrac{d\varphi}{d \tau} = L$ (for photon we have $\tau=\lambda$). Since the metric is static we have an energy conservation $E=A(r)\dfrac{dt}{d\tau}$
(for photon we have $\tau=\lambda$). Assume that $A(r) > 0$.

\begin{figure}[!tbh]
\centering
\includegraphics[width=0.8\linewidth]{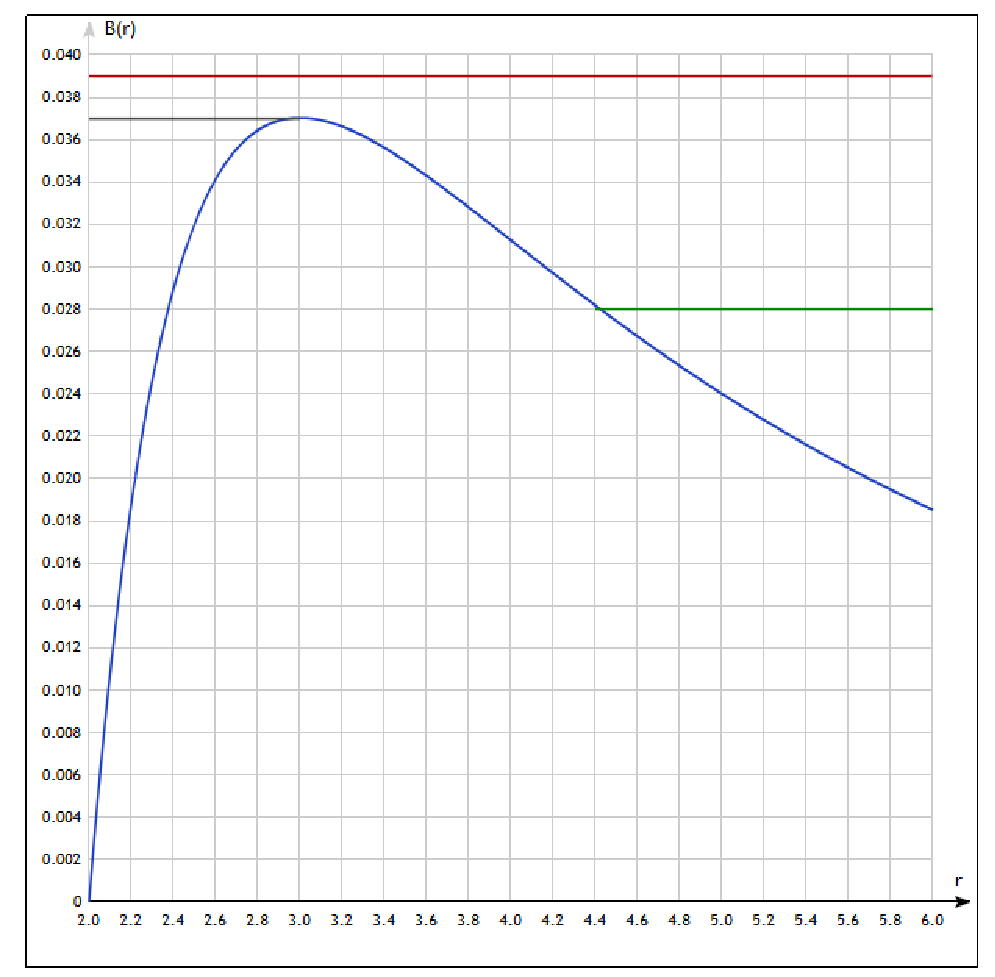}
\caption{Blue curve represents $(1-2/r)/r^2$ for all $r$. Function $B(r)$ corresponds to the blue curve for  $r \geq 3$
and corresponds to the black horizontal straight line $B(r)=1/27$ for  $r < 3$.  Red horizontal straight line $b=5.063$ ($B(r)=0.039$) corresponds to capture of photon,
the green horizontal straight line $b=6$ ($B(r)=0.028$ for $r> 4.4$) corresponds to the case when photon moves from infinity  toward the  until $r \approx 4.4$, at this point photon turns and moves to infinity again. Critical impact parameter $b=3\sqrt{3}$ separates scatter and capture regions for the impact parameters.
It is the same as for the Schwarzschild metric.
 }\label{fig:plot1}
\end{figure}

Let us test the metric done in Eq. (\ref{M_0}) with photons ($ds^{2}=0$). We consider this case just for simplicity but it is possible to use the same analysis for massive particles moving
in the metric. Then, we obtain from Eq. (\ref{M_0}) 
\begin {equation}
A(r)\left(\dfrac{dt}{d\lambda}\right)^{2}-A^{-1}(r)\left(\frac{dr}{d\lambda}\right)^{2}+
r^{2}\left( \dfrac{d \phi}{d\lambda}\right)^{2}=0,
\label{M_1}
\end {equation}
since
\begin {equation}
\left(\dfrac{dt}{d\lambda}\right)^{2}=\dfrac{E^2}{A^2(r)},
\label{M_2}
\end {equation}
we obtain
\begin {equation}
\left(\dfrac{dr}{d\lambda}\right)^{2}=\left(\dfrac{E^2}{A(r)}-\dfrac{L^2}{r^2}\right) A(r),
\label{M_3}
\end {equation}
or
\begin {equation}
\left(\dfrac{dr}{d\lambda}\right)^{2}=\left(\dfrac{1}{A(r)}-\dfrac{b^2}{r^2}\right) \frac{ A(r)}{E^2},
\label{M_4}
\end {equation}
where $b=L/E$. Therefore, the photon motion can only in regions where
\begin {equation}
\dfrac{1}{A(r)}\geq \dfrac{b^2}{r^2},
\label{M_5}
\end {equation}
or
\begin {equation}
B(r) \leq \dfrac{1}{b^2},
\label{M_6}
\end {equation}
where $B(r)=A(r)/r^2$.
If we define function $A(r)= 1-2/r$ for $r \geq 3$, while $A(r)=r^2/27$ for $r < 3$ (therefore, $B(r)= (1-2/r)/r^2$ for  $r \geq 3$ and  $B(r)=1/27$ for $r < 3$ . 
As we noted above we could compare Eq. (6) from Friedman paper \cite{Friedman_22}  and Eqs. (\ref{M_3}) or  (\ref{M_4}) and our Fig. \ref{fig:plot1}
with Figure in \cite{Friedman_22}. It is not strange since in both papers the authors were searching regions where some functions are positive.

\begin{figure}[!tbh]
\centering
\includegraphics[width=0.8\linewidth]{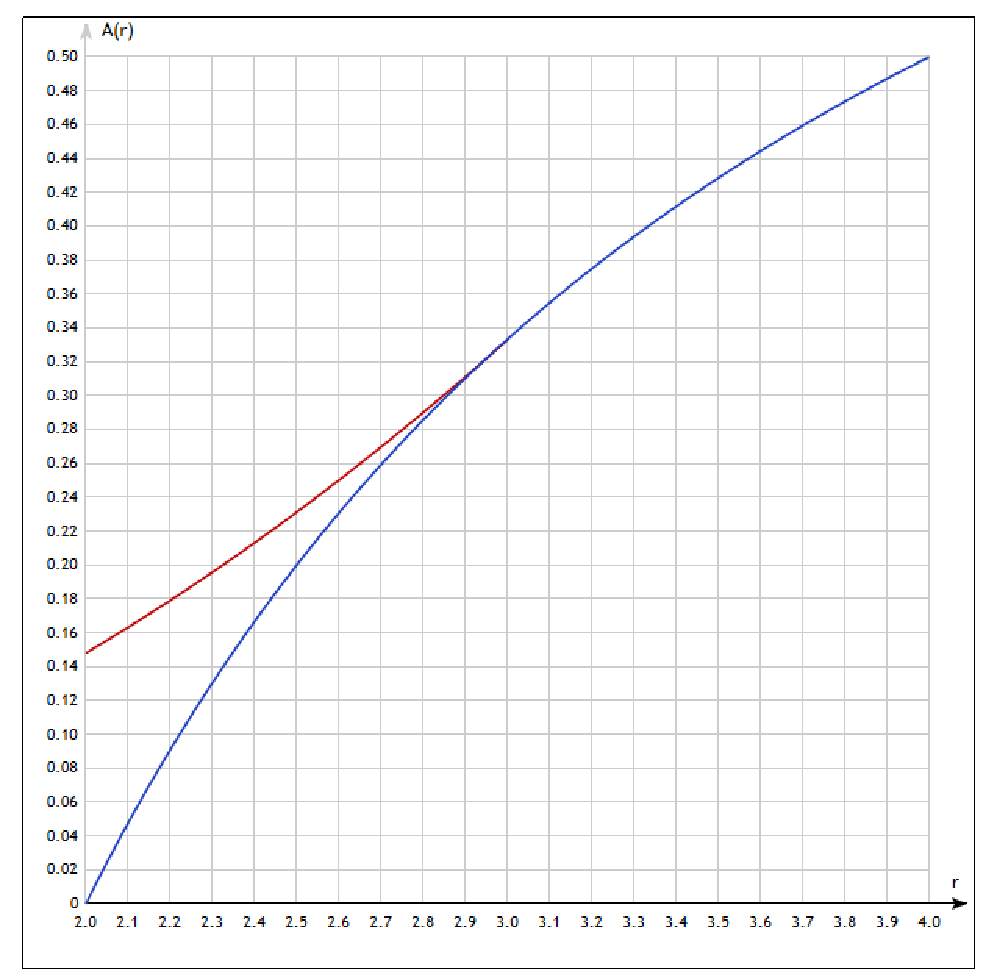}
\caption{Blue curve presents function $A(r)$  for the Schwarzschild, while red curve corresponds to  our modified $A(r)$. These red and blue curves are different only
for $r < 3$.}\label{fig:plot2}
\end{figure}

If we conduct a thought experiment like Rutherford's experiment on the scattering of photons by a Schwarzschild black hole, we will not be able, even in principle, to distinguish the metric of a black hole from a metric where the function $A(r)$ is replaced by a modified one  in the interval $2 < r < 3$.
Similarly, if we use bound orbits to test the black hole metric, the Schwarzschild metric will be indistinguishable 
from the modified metric considered above and done in Eq. (\label{M_1}) , 
since the probe body tests gravity only in the range of radial coordinate 
periapsis to apoapsis but in this region the metric in Eq. (\label{M_1}) coincides with Schwarzschild one.
We have considered only two ways to test the metric of a compact object, but even when considering other approaches there are limitations to the inference that the processing of observations leads to the conclusion that there is a Schwarzschild and/or Kerr black hole. It is probably more correct in this case to say that this black hole model describes the observational data best among the alternatives considered.

\section{Conclusions}

In this article we recalled some fragments of the development of physics and astronomy in Russia, not all of them are well known even to specialists. The remarkable words of  S.~Weinberg about the history of physics come to mind. 
At the end of his life this famous scientist worked in GR and astronomy and wrote several outstanding books on these subjects \cite{Weinberg_72,Weinberg_93,Weinberg_08,Weinberg_20}.
In one of his last popular book Weinberg emphasized an importance of history of science and wrote:''I am a physicist, not a historian, but over the years I have become increasingly fascinated by the history of science. It is an extraordinary story, one of the most interesting in human history. It is also a story in
which scientists like myself have a personal stake. Today’s research can be aided and illuminated by a
knowledge of its past, and for some scientists knowledge of the history of science helps to motivate
present work'' \cite{Weinberg_15}. 

Unfortunately, in Russia the achievements of domestic scientists were sometimes hushed up, thereby giving additional bonuses to foreign scientists.
In this context, one can recall the underestimation of the importance of Friedman's work on cosmology, since Soviet philosophers and ideologists insisted that the Universe is infinite in time and space. On May 2, 1946, P. L. Kapitsa wrote a letter to Stalin, asking for support for the publication of Gumilevsky's book ''Russian Engineers'',
in particular, Kapitsa wrote \cite{Kapitsa_46} ''it is clear from Gumilevsky's book that
a large number of the largest engineering initiatives were born in our country, we ourselves were almost unable to develop them, often 
often the failure to use innovation is that we usually underestimated our own and overestimated foreign...Many of the organizational shortcomings still exist today, and one of the main ones is the underestimation of one's own and overestimation of foreign forces....After all, excessive modesty is an even greater disadvantage than excessive self-confidence...We will do this successfully [to develop national technology] only when we believe in the talent of our scientist and engineer...''
A number of measures were taken that led to the acceleration of the country's technological development, the social status of scientists was raised, but such ugly campaigns as the fight against sycophancy and cosmopolitanism also arose. Perhaps Kapitsa's letter to Stalin influenced his determination to begin the fight against servility to the West.
Soviet writer and poet K. M. Simonov recalled that one at the meeting with writers Stalin said:
''There is an issue which is very important which writers need to be interested in. This is the theme of Soviet patriotism. If we take our average intelligentsia, scientific intelligentsia, professors, doctors, they have not developed a sufficient sense of conscientious patriotism. They have an unjustified admiration for foreign culture. They all feel like minors, not one hundred percent, they are used to considering themselves in the position of eternal students... First the Germans, then the French, there was admiration for foreigner'' \cite{Simonov_88}.

As an example of the disgusting campaign to combat cosmopolitanism and servility in 1940s and 1950s, one can recall one of the days of V. L. Ginzburg, who later became the last Russian Nobel laureate in physics.
On Ginzburg's birthday (on October 4, 1947), Soviet Literaturnaya Gazeta (Literature Newspaper) published an article by economist V. Nemchinov with a headline ''Against servility!'' \cite{Nemchinov_47}. The full text of this article by Nemchinov can be found also in the anniversary album dedicated to the 100th anniversary of Ginzburg                  \cite{Ginzburg_17b}\footnote{V. Nemchinov was an academician,  one of the leaders among Soviet economists and statisticians.
He promoted mathematical methods in economy and we was
  the  Rector of the Timiryazev Agricultural Academy (TSKhA) in 1947.
Since he  defended genetics and TSKhA scientists working in this direction of biology   at the August Session in 1948 of the 
Lenin All-Union Academy of Agricultural Sciences (VASKhNIL) \cite{Shnol_10}, soon after the session he was resigned from the rector position. 
See, also, a detailed report on the event in ''On the Situation in Biological Science.
Verbal Report Session of the All-Union Academy of Agricultural Sciences named after V. I. Lenin'' \cite{VASKhNIL_48}, practically, it was a defeat of genetics, after which genetics could not recover for several decades.
As it was noted by P. Pringle at the session it was not mentioned not once the VASKhNIL founder and first president  N. I. Vavilov, nor the biological institutes that Vavilov founded and led \cite{Pringle_08} (N. I. Vavilov died in prison in 1943).}.
In the Nemchinov's article it was 
noted that in his brochure on the atomic nucleus doctor Ginzburg does not refer to the Iwanenko-Gapon model of the nucleus, the author of the article writes that the author is hushing up this achievement of Soviet science.
As for the lack of citation of the Iwanenko--Pomeranchuk article on synchrotron radiation, it is said that there is a completely absurd groveling before American science, since Ginzburg cited a review by an American author. There is nowhere to go beyond this shameful phenomenon of hushing up discoveries of Soviet science, of erasing Soviet authors.
Once at the house of mutual friends Ginzburg met a person (Schneiderman) who was the real author of the article signed by V.~Nemchinov and in these conversations Schneiderman said that authorities requested him
to write an article criticizing Lysenko opponents, basically genetics, including A. R. Zhebrak. Ivanenko worked in  TSKhA, learned about the article being prepared
 and recommended to include
cases from other branches of science to reach a greater generality. 
So, V. L. Ginzburg's was sure that the Nemchinov's article was initiated by D. D. Iwanenko, however, in 1947 V. L. Ginzburg started to work in the Soviet Atomic 
project\footnote{Ginzburg's key contribution to the creation of the hydrogen bomb is described in an article by one of the leaders of the Soviet atomic project, Yu. B. Khariton
\cite{Khariton_96,Khariton_99} (see also \cite{Goncharov_96,Goncharov_97,Goncharov_05}).} and  according to his opinion, only these circumstances influenced the fact that he was not subjected to repression, since he was married to a repressed woman, as a cosmopolitan and a Jew \cite{Ginzburg_01}. On the same day, October 4, 1947, the plenum of the Higher Attestation Commission decided not to approve Ginzburg's academic title of professor \cite{Ginzburg_01} (two ''gifts'' for one birthday).

We outlined individual stages of development of research in the theory of gravity, in particular, in cosmology.
In September of this year it will be possible to sum up the results of 100 years of development of physical cosmology after the departure of A. A. Friedman,
and in November it will be time to sum up the results of development of GR for 110 years after its creation by A. Einstein. 

Coming to the conclusion of the discussion of tests of general relativity, in particular, the new test associated with the reconstruction of shadows of black holes from observations, it can be noted that several years have passed since the reconstruction of the black hole shadow in M87* and the Galactic Center was presented. The images of the rings around these shadows have sometimes even been called images of black holes. However, the main thing in these images is the dark region inside the rings, since it is this region that characterizes the gravity of the black hole. One of leader EHT Collaboration wrote a popular book \cite{Falcke_21} (in Russian edition of the book
in contrast to the picture at the envelope of the book in German, English and Italian editions, however, in Russian the edition printed in 2024,
publishers changed the EHT ring image for M87* with an artist view picture which is wrongly reflects a gravitational field near a black hole,
therefore they removed the main essence of the book and the author activity for decades)
and paper \cite{Falcke_22} describing the way to reconstruct a bright ring around darkness.

We described how pure theoretical shadow concept was transformed in the observational quantity which can be obtained from observations as the EHT collaboration
demonstrated. As R. Genzel concluded in his Nobel prize lecture any plausible astronomical model for the Galactic Center must include 
a supermassive black hole with a mass around $4\times 10^{6}M_\odot$ \cite{Genzel_22}.

\section*{Acknowledgements}

This paper was based on the talk presented at the Rubakov-70 conference. The author appreciates
participants of the conference for fruitful conversations which helped me to clarify 
some issues and to form a historical part of the paper. 
The author also thanks professor V. O. Soloviev for useful discussions of history of physics and related issues.
The author
appreciates also V. A. Matveev for his invitation to submit the paper in the Natural Science Review electronic journal. 

\section*{Funding}
This  research  was supported by ongoing institutional funding.
No additional grants to carry out or direct this particular
research were obtained.

\section*{Conflicts of Interest}

The author declares no conflicts of interest.

\bibliographystyle{nsr}
\bibliography{Zakharov_biblio}

\end{document}